\documentclass[reprint,amsmath,amssymb,aps,prb]{revtex4-2}

\usepackage[utf8]{inputenc}
\usepackage{graphicx}
\usepackage{xcolor}
\usepackage{gensymb}
\usepackage{subcaption}
\usepackage{bm}

\begin{document}
\makeatletter
\def\frontmatter@date@produce{}
\makeatother

\title{Effect of the Gradient of the Spin-Polarization in Density Functional Approximations}

\author{Rohan Maniar}
\author{John P. Perdew}
\affiliation{Department of Physics and Engineering Physics, Tulane University, New Orleans, LA 70118, USA}

\maketitle

\section*{Abstract}

The construction of non-empirical density functional approximations is typically guided by the satisfaction of exact constraints. An important constraint is the recovery of the gradient expansion for slowly varying electron densities. In prior constructions of semilocal density functional approximations, the $\nabla \zeta$-dependent terms in the gradient expansion of the correlation have been dropped, where $\zeta$ is the relative spin polarization. We propose a scheme by which such terms can be reintroduced into already constructed functionals without significantly affecting other constraints and norms. We implement this scheme on the Strongly Constrained and Appropriately Normed (SCAN) functional to construct a $\nabla \zeta$-corrected version of SCAN. The resulting functional is shown to provide improvements in transition-metal atoms and molecules without significantly affecting SCAN's accurate description of $sp$-systems. For the binding energy curve of the chromium dimer Cr$_2$, the SCAN  underbinding is fully corected at large bond lengths and reduced at short bond lengths.

\section{Introduction}

Kohn-Sham density functional theory \cite{hohenberg1964inhomogeneous,kohn1965self} provides the theoretical framework for most electronic structure calculations performed today. While exact in principle, one always has to approximate the exchange-correlation functional ( $E_{\mathrm{xc}}[n_{\uparrow},n_{\downarrow}]$ ) in practical calculations. Local and semilocal approximations for $E_{\mathrm{xc}}[n_{\uparrow},n_{\downarrow}]$ have become particularly popular as they involve single integrals and are hence computationally attractive, especially for larger systems. Such approximations take the general form:

\begin{equation}
    E_{\mathrm{xc}}[n_{\uparrow},n_{\downarrow}] = \int d^{3}r \ n({\mathbf{r}}) \ \epsilon_{\mathrm{xc}}([n_{\uparrow},n_{\downarrow}] ; \mathbf{r}) ,
\label{Exchange-correlation form}
\end{equation}

\noindent
where $\epsilon_{\mathrm{xc}}([n_{\uparrow},n_{\downarrow}]; \mathbf{r})$, the exchange-correlation energy per particle, usually depends on the local density( $n_{\uparrow}(\mathbf{r}),n_{\downarrow}(\mathbf{r})$ ),  its gradient ($\nabla n_{\uparrow}(\mathbf{r}),\ \nabla n_{\downarrow}(\mathbf{r})$), and the non-interacting kinetic-energy density ($\tau_{\uparrow}(\mathbf{r}),\ \tau_{\downarrow}(\mathbf{r})$). he total density is defined as $n(\mathbf{r})=\sum\limits_{\sigma} n_{\sigma}(\mathbf{r})$.

A useful construction principle for these semi-local approximations is the satisfaction of exact constraints \cite{kaplan2023predictive}. Evidence for this is the SCAN functional \cite{sun2015strongly}, which satisfies all 17 exact constraints known for a semi-local functional,  and is remarkably accurate for both molecules and solids \cite{sun2016accurate,isaacs2018performance}. 

A semilocal ingredient, dropped in SCAN's correlation but retained implicitly in its exchange, is the gradient $\nabla \zeta$ of the relative spin polarization, $\zeta=\frac{n_{\uparrow}-n_{\downarrow}}{n_{\uparrow}+n_{\downarrow}}$. A recent investigation has shown that $\nabla \zeta$-dependent  terms in the correlation energy, when introduced within the spin-current DFT framework, are potentially important \cite{desmarais2025meta}. In this work, we show that the $\nabla \zeta$-dependent gradient expansion terms, derived for the high-density limit, when added to SCAN's usual correlation, can produce improvements, especially for transition metal systems. Furthermore, we used the new semilocal ingredient ($\nabla \zeta$) to cancel the residual $\zeta$ dependency in the low-density correlation of SCAN. 

The original $\nabla \zeta$-dependent gradient expansion terms in the correlation were derived by Rasolt and collaborators in the high-density limit (HDL) \cite{rasolt1977inhomogeneity,rasolt1981exchange}. These terms were later parameterized by Perdew et al. \cite{perdew1992atoms,wang1991spin} as follows:

\begin{widetext}
\begin{equation}
\begin{split}
    \Delta E_c^{\mathrm{HDL}}\left[n_{\uparrow}, n_{\downarrow}\right] 
    \approx \mathrm{C_c(0)} \int d^3 r n \left\{ 
    \frac{-0.458 \zeta \nabla \zeta \cdot}{\left[n\left(1-\zeta^2\right)\right]^{1 / 3}} \left(\frac{\nabla n}{n}\right) 
    + \frac{\left(-0.037+0.10 \zeta^2\right)|\nabla \zeta|^2}{n^{1 / 3}\left(1-\zeta^2\right)}
    \right\},
\end{split}
\label{High-density correction to correlation}
\end{equation}
\end{widetext}

\noindent
in atomic unit, where \( \mathrm{C_c(r_{s})} \) is a function of \( r_{s} = \left( \frac{3}{4 \pi n} \right)^{\frac{1}{3}} \), and \( \mathrm{C_c(0)} = 0.004235 \). These terms tend to, but do not necessarily, provide positive contributions to the total energy for spin-polarized systems \cite{perdew1992atoms}. While usually small for $sp$ atoms, when added to the Perdew-Wang GGA (PW92) functional \cite{wang1991spin}, these correlation terms have been shown to provide non-negligible contributions for transition metal atoms. For example, for the Cr atom, the $\nabla \zeta$-dependent term pushed the PW92 total energy up by 0.62 eV \cite{perdew1992atoms}. It hence becomes important to add the terms in Equation \ref{High-density correction to correlation} to SCAN's usual correlation in the HDL. We will "turn off" this contribution to the correlation slowly as we move away from the HDL (r$_{s} \rightarrow 0$).

In the opposite low-density limit (LDL), for a slowly-varying electron gas, we identify that SCAN deviates from an exact constraint- the weak dependence on the spin polarization- owing to the omission of $\nabla \zeta$ terms in its correlation. Note that SCAN implicitly retains such $\nabla \zeta$-dependent terms in its exchange part, which, in the low-density limit, must be canceled by counter-acting terms in the correlation. 

Starting from the gradient expansion for a spin-unpolarized gas, terminated at second order in the reduced density gradient $s=\frac{|\nabla n |}{2 (3 \pi^{2})^{\frac{1}{3}}n^{\frac{4}{3}} }$\cite{antoniewicz1985kohn,svendsen1995gradient,svendsen1996gradient}:
\begin{equation}
E_{\mathrm{x}}[n]=A_{\mathrm{x}} \int \mathrm{~d}^3 r n^{4 / 3}\left[1+\mu s^2+\ldots\right],
\end{equation}

\noindent
and using the spin-scaling relations \cite{oliver1979spin} for the exchange ($E_{x}[n_{\uparrow},n_{\downarrow}] = \frac{1}{2}E_{x}[2n_{\uparrow}]+\frac{1}{2}E_{x}[2n_{\downarrow}]$), the required correction to the correlation in the LDL (r$_{s} \rightarrow \infty$) can be derived:

\begin{widetext}
\begin{equation}
\Delta E_c^{\mathrm{LDL}}\left[n_{\uparrow}, n_{\downarrow}\right] 
\approx \mathrm{-C_c(\infty)} \int d^3 r n \left\{ \frac{\nabla \zeta \cdot \nabla n}{n} \left(\frac{1}{n_{\uparrow}^{\frac{1}{3}}} - \frac{1}{n_{\downarrow}^{\frac{1}{3}}}\right) 
+ \frac{n |\nabla \zeta|^2 }{4} \left(\frac{1}{n_{\uparrow}^{\frac{4}{3}}} + \frac{1}{n_{\downarrow}^{\frac{4}{3}}}\right) \right\},
\label{Low-density correction to correlation}
\end{equation}
\end{widetext}

\noindent
where $\mathrm{C_{c}(\infty)} = \frac{\mu A_{x}}{8} (\frac{2}{3\pi^{2}})^\frac{2}{3}$, $\mu=\frac{10}{81}$, and $A_{x}=-\frac{3}{4 \pi} (3 \pi^{2})^{\frac{1}{3}}$. In Equation \ref{Low-density correction to correlation}, n refers to the total spin density, and $n_{\sigma}$ refers to its spin-up and spin-down contributions. As for the HDL correction, terms in Eq.\ref{Low-density correction to correlation} will be "turned off" as we move to higher densities.

To apply the appropriate $\nabla \zeta$ correction terms in the two discussed limit, we define a controlled interpolation between the high-density and low-density limits  as follows:

\begin{equation}
\Delta E_c \mathrm \ {=} \ \Delta E_{c}^{\mathrm{HDL}} \textit{e}^{\beta_{1} (1-(1+r_{s}^{2})^{\frac{1}{4}})} \ + \ \Delta E_{c}^{\mathrm{LDL}} (1 - \textit{e}^{\beta_{2} (1-(1+r_{s}^{2})^{\frac{1}{4}})}),
\label{Intepolation of corrections}
\end{equation}

\noindent
where $r_{s}$ is the Seitz radius, and two parameters, which are not necessarily equal- $\beta_{1}$ and $\beta_{2}$- are introduced, to be determined via fitting to appropriate norms. Furthermore, the  $\nabla \zeta$ correction must be damped whenever its contribution is large in magnitude, compared to SCAN's original correlation, in the slowly-varying limit (where the iso-orbital indicator, $\alpha=\frac{\tau-\frac{|\nabla n|^{2}}{8n}}{\frac{3}{10} (3 \pi^{2})^{\frac{2}{3}}n^{\frac{5}{3}}}$, approaches 1). This gives the following form for the correlation in the slowly-varying limit:

\begin{subequations}
\begin{equation}
    \epsilon_{c} ( \alpha = 1) = \epsilon_{c}^{\mathrm{SCAN}} ({\alpha =1}) \Big( 1+\frac{v}{1+v^{2}} \Big)
\end{equation}
\begin{equation}
    v=\frac{\Delta E_{c}}{\epsilon_{c}^{\mathrm{SCAN}}(\alpha = 1)} 
\end{equation}
\label{gcz-SCAN slowly-varying limit}
\end{subequations}

\noindent
where $\epsilon_{c}^{\mathrm{SCAN}}$($\alpha$=1) is the original generalized gradient approximation used by SCAN in this limit.

Using this new correlation energy per particle for the slowly-varying limit ($\alpha$=1), and retaining SCAN's iso-orbital ($\alpha$=0) description, correlation interpolation ($f_{c}(\alpha)$) , and exchange energy per particle ($\epsilon_{x}^{\mathrm{SCAN}}$), gives us a new functional that we call gzc-SCAN (gradient-zeta-corrected SCAN). The parameters $\beta_{1} (=0.240)$ and $\beta_{2}(=0.033)$, introduced in the functional form, are determined by minimizing the mean absolute percentage errors for the non-relativistic total atomic energies of Li and Na \cite{chakravorty1993ground}. Note that, as all of SCAN's norms involved either spin-unpolarized or fully-spin-polarized systems, none of SCAN's original parameters require readjustment.

To the extent that Eq.6a makes a positive contribution to the exchange-correlation energy density of SCAN, the generalized Lieb-Oxford \cite{perdew2014gedanken,lieb1981improved} lower bound will be satisfied. Except for this bound,the gzc-SCAN functional satisfies all constraints that its parent functional, SCAN, does.This includes the non-uniform density scaling relation \cite{levy1991density}, which is preserved as we do not vary SCAN's iso-orbital ($\alpha$=0) description \cite{perdew2014gedanken}.  gzc-SCAN is hence expected to mostly retain SCAN's high accuracy for atoms, molecules, and solids. This is verified by its decent performance for atomization energies and barrier heights (although not as accurate as SCAN itself). In addition, the gzc-SCAN functional gives an improved description of transition-metal systems, including ionization energies of transition metal atoms and more accurate binding energy curves for transition-metal dimers.  

The proposed scheme to include $\nabla \zeta$-dependent terms can be implemented for any GGA, or any meta-GGA that can identify the slowly-varying-density limit through an iso-orbital indicator.

\section{Computational Details}

We restrict our current investigation of gzc-SCAN to atoms and molecules. The functional has been implemented in the all-electron UTEP-NRLMOL code \cite{porezag1999optimization,pederson1990variational}, which features a highly accurate numerical integration grid and extensive basis sets. Integer occupations were ensured, and an energy tolerance of $10^{-6}$ Hartree was used for all calculations.  The standard NRLMOL basis set is supplemented with Gaussians at the bond center for the Cr dimer calculations, as done in Ref \cite{maniar2024symmetry}.

The functional has also been implemented within the projector-augmented wave (PAW) method \cite{blochl1994projector,kresse1999ultrasoft} in the Vienna \textit{Ab initio} Simulation Package (VASP) \cite{kresse1993ab,kresse1994ab,kresse1996efficiency,kresse1996efficient}. In the specific case of the Mn dimer, we use the VASP implementation of the functional, for an easier comparison with Ref \cite{lopez2025revisiting} . An energy cutoff of 800 eV was used, and the pseudopotentials were those constructed for 3s and 3p as well as 3d and 4s valence electrons. Sufficiently large cells ($10~\text{\AA} \times 10.1~\text{\AA} \times 13.5~\text{\AA}$) were chosen to avoid interactions with periodic images.

We mention in passing that due to the presence of $\nabla \zeta$-dependent terms in the gzc-SCAN correlation, it does not depend on $|\nabla n|$ but on $\nabla n_{\sigma}$. The generalized-Kohn-Sham potential, $v_{xc}^{\sigma} (\mathbf{r})$, is consequently constructed as \cite{jana2018assessing}:

\begin{equation}
v_{x c}^{\sigma} \Psi_{i\sigma}=\left[\frac{\partial\left(n \epsilon_{x c}\right)}{\partial n_{\sigma}}-\vec{\nabla} \cdot \frac{\partial\left(n \epsilon_{x c}\right)}{\partial \vec{\nabla} n_{\sigma}}\right] \Psi_{i\sigma}-\frac{1}{2} \vec{\nabla}\left(\frac{\partial\left(n \epsilon_{x c}\right)}{\partial \tau_{\sigma}}\right) \cdot \vec{\nabla} \Psi_{i\sigma} ,
\end{equation}

\noindent
where the $\psi_{i \sigma}$ represent the generalized-Kohn-Sham orbitals.

All calculations (in both NRLMOL and VASP) have been performed using fixed geometries.
\section{Results}

\subsection{Total Energies of Atoms}
We begin this section by investigating the effect of the $\nabla \zeta$-dependent correction terms on the total energies of atoms. One may expect large total energy contributions from such terms to come from regions where one transitions from the closed-shell core ($\zeta$=0) to the spin-polarized valence ($\zeta\neq$0). This effect should be particularly pronounced in transition metal systems, which involve highly localized spin-polarized $ d$- and $ f$-shells. 

To test this expectation, we look at the difference between SCAN and gzc-SCAN total energies for the $M^{+3}$ cations of the first-row transition metal atoms ($M$= Sc-Zn). The $M^{+3}$ cations were chosen as they do not involve $4s$ electrons in their electronic configuration, avoiding complications due to potential $3d$-$4s$ mixing. From Table \ref{Total energy difference atoms}, one sees a monotonic increase in the difference between the total energies of gzc-SCAN and SCAN total energies as the number of unpaired $ 3d-$ electrons increases. The total energy differences can be as high as 0.44 eV, as found for the case of $\mathrm{Fe^{+3}}$. These potentially large shifts in total energies through the addition of $\nabla \zeta$-dependent terms become particularly relevant in our discussion of transition metal systems. 

\begin{table}[t]
\caption{ Difference between gzc-SCAN and SCAN total energies ($\Delta$) as a function of the number of unpaired 3d electrons ($N_{\mathrm{unpaired}}$) for the $M^{+3}$ cations, where $M$=Sc-Zn. The $\nabla \zeta$ terms added to the correlation tend to make the energy of the spin-polarized atoms more positive. The total energies are reported in atomic units, and the energy differences are reported in eV.} 
\centering
\begin{tabular}{  c  c  c  c  c  c }
\hline
\hline
 $M$ & $N_{\mathrm{unpaired}}$ & SCAN & gzc-SCAN & $\Delta$ (eV) \\
 \hline
Sc & 0 & -759.0477 & -759.0477 & 0.00 \\
Ti & 1 & -847.6516 & -847.6501 & 0.04  \\
V &  2 & -942.1116 & -942.1084 & 0.09 \\
Cr &  3 & -1042.5201 &-1042.5132 & 0.19 \\
Mn   & 4 & -1149.0085 & -1148.9971 & 0.31 \\
Fe  & 5 & -1261.7641 & -1261.7478 & 0.44 \\ 
Co & 4 & -1380.6955 &  -1380.6842 & 0.31 \\
Ni & 3 & -1506.1705 & -1506.1634 & 0.19\\
Cu & 2 & -1638.2714 & -1638.2674 & 0.11 \\
Zn & 1 & -1777.1072 & -1777.1062 & 0.03 \\
\hline
\hline
\end{tabular}

\label{Total energy difference atoms}
\end{table}

\subsection{Atomization Energies and Barrier Heights for sp molecules}

Before further testing gzc-SCAN for transition metal systems, it is essential to first check that the functional retains SCAN's good performance for $sp$ molecules \cite{sun2016accurate,dasgupta2021elevating}. We begin by reporting errors for small test sets \cite{lynch2003small}  that are representative of errors in the atomization energies (AE6) and barrier heights (BH6) of the much larger Database/3 data set \cite{lynch2003robust}. These errors are compared with those of PBE \cite{perdew1996generalized}, TPSS \cite{tao2003climbing}, and SCAN \cite{sun2015strongly}. 

From Table \ref{tab: AE6 and BH6 errors}, one finds that the addition of the $\nabla \zeta$ terms to SCAN's correlation appears to deteriorate its performance for $sp$ molecules, but no worse than that of TPSS. We point out that the positive shift in SCAN's mean error (by 4.8 kcal/mol) for the AE6 dataset, on addition of $\nabla \zeta$ terms to its correlation, is because such terms push the energy of the spin-polarized atoms in the dataset upwards (make them less negative) while leaving the energies of the closed-shell molecules unchanged. As the TPSS mean error is already positive, it is likely that recovering the $\nabla \zeta$ terms in its slowly-varying limit will worsen its performance for the AE6 dataset for the same reason. The BH6 errors are not affected very much by the addition of the $\nabla \zeta$ terms, as, for barrier heights, delocalization errors play a more important role \cite{kaplan2023understanding,singh2024rise}.

\begin{table}[h]
	\begin{subtable}[h]{0.49\textwidth}
		\centering
  		\caption{AE6}
		\begin{tabular}{l  l  l}
        \hline
        \hline
		  Functional & ME & MAE\\
        \hline
		PBE & 10.6 & 13.8 \\
		TPSS &  4.1  & 5.9  \\
		SCAN & 0.2 & 3.0 \\
		gzc-SCAN & 5.1 & 5.9\\
        \hline
        \hline
		\end{tabular}

		\label{tab: AE6 errors}
	\end{subtable}
	\hfill
	\begin{subtable}[h]{0.49\textwidth}
		\centering
  	\caption{BH6}
		\begin{tabular}{ l l l }
        \hline
        \hline
		  Functional & ME & MAE\\
		\hline
		PBE & -9.6 &  9.6 \\
		TPSS &  -8.3 &  8.3 \\
		SCAN & -7.9 & 7.9\\
		gzc-SCAN & -8.4 & 8.4 \\
        \hline
        \hline
		\end{tabular}
	
		\label{tab: BH6 errors}
	\end{subtable}
    	\caption{ gzc-SCAN errors for the AE6 and BH6 datasets compared with those of PBE, TPSS, and SCAN. TPSS errors for AE6 and BH6 were taken from Ref \cite{tao2017semilocal}. All errors are reported in kcal/mol (1 kcal/mol $=$ 0.0434 eV). }
	\label{tab: AE6 and BH6 errors}
\end{table}

We now move to reporting errors in the larger G3-1 \cite{curtiss2000assessment} and BH76 \cite{zhao2005benchmark} test sets in Table \ref{tab: G3-1 and BH76}. The G3-1 test set benchmarks enthalpies of formation whose errors are equal in magnitude and opposite in sign to the error in the atomization energy. For the G3-1 test set, SCAN and gzc-SCAN produce similar MAEs, however, the ME indicate the expected tendency of gzc-SCAN to overbind constituent atoms in molecules when compared with SCAN. We suspect that these overbinding tendencies will worsen errors in the predicted enthalpies of formations in the G3-2 and G3-3 test sets which contain larger molecules and where SCAN already overbinds. However, as noted in Ref \cite{perdew2016density}, atomization energies tend to magnify small errors in spin-polarization energies of atoms that may not be important for other properties of molecules and solids. BH76 errors closely resemble those of BH6. 

\begin{table}[h]
	\begin{subtable}[h]{0.49\textwidth}
		\centering
  		\caption{G3-1}
		\begin{tabular}{ l l l }
        \hline
        \hline
		  Functional & ME & MAE\\
		\hline
		PBE & -5.3 & 7.0 \\
		TPSS &  -3.7  & 4.5  \\
		SCAN & 1.0 & 3.5 \\
		gzc-SCAN & -2.0 & 3.2\\
        \hline
        \hline
		\end{tabular}

		\label{tab: G3-1 errors}
	\end{subtable}
	\hfill
	\begin{subtable}[h]{0.49\textwidth}
		\centering
  	\caption{BH76}
		\begin{tabular}{ l l l }
        \hline
        \hline
		  Functional & ME & MAE\\
		\hline
		PBE & -9.3 &  9.3 \\
		TPSS &  -8.6 &  8.7 \\
		SCAN & -7.7 & 7.8\\
		gzc-SCAN & -8.0 & 8.1 \\
        \hline
        \hline
		\end{tabular}
	
		\label{tab: BH76 errors}
	\end{subtable}
    	\caption{ gzc-SCAN errors for the G3-1 and BH76 datasets compared with those of PBE, TPSS, and SCAN. TPSS errors for the G3-1 and BH76 test sets were taken from Ref \cite{sun2015strongly}. All errors are reported in kcal/mol. Note that the error of the G3-1
     formation energy is opposite in sign to the error of the AE6 atomization energy.} 
	\label{tab: G3-1 and BH76}
\end{table}

\subsection{Ionization Energy of Transition Metal Atoms}

Ionization energies are an easy test for DFAs owing to their ease of computation and the abundance of available experimental data \cite{NIST_ASD}. It is hence natural, as a preliminary test for transition metal systems, to check the performance of gzc-SCAN for the ionization energies of first-row transition metal atoms (Sc-Zn).

Recent work has shown that, for certain atoms/cations considered, the electronic configurations obtained from DFT calculations do not necessarily match those experimentally determined \cite{maniar2025atomic}. In these cases, Ref \cite{maniar2025atomic} re-evaluated the reference ionization energies to be consistent with DFA electronic configurations and showed such reference values to be more revealing of deficiencies in DFAs. In this work, we use the reference ionization energy values, consistent with r$^{2}$SCAN's electronic configurations, as reported in Table S2 of Ref \cite{maniar2025atomic}, when evaluating gzc-SCAN errors (shown in Table \ref{TM IP errors}). 

In Table \ref{TM IP errors}, we find the expected trend of increasing accuracy as we climb up Jacob's ladder. We also note that gzc-SCAN provides the lowest mean absolute errors of all functionals considered, even smaller than those obtained from r$^{2}$SCAN, for 
the first three ionization energies.

\begin{table}[t]
\centering
\begin{tabular}{  c c  c  c  c  c c  c  c  c  c }
\hline
\hline
  &  1$^{\mathrm{st}}$ IP  &  2$^{\mathrm{nd}}$ IP & 3$^{\mathrm{rd}}$ IP \\
\hline
LSDA & 0.53 & 1.06 & 1.11 \\
PBE & 0.45 &  0.72 &  0.77 \\
r$^{2}$SCAN & 0.42 &  0.35 &  0.39\\
gzc-SCAN & 0.33 & 0.33 & 0.32 \\
\hline
\hline
\end{tabular}

\caption{ Mean absolute errors for the first three ionization energies of first-row transition metal atoms from various DFAs. The errors for LSDA, PBE, and r$^{2}$SCAN (Ref \cite{maniar2025atomic} does not report SCAN errors), which are evaluated using appropriately adjusted reference values, were taken from Ref \cite{maniar2025atomic}. All errors are reported in eV.} 

\label{TM IP errors}
\end{table}

\subsection{Transition Metal Dimers}

We finally briefly look at the chromium (Cr$_{2}$) and manganese (Mn$_{2}$) dimers using the gzc-SCAN functional. Both dimers are challenging for standard density functional approximations \cite{maniar2024symmetry,ivanov2021mn}, with the Cr$_{2}$ dimer even being branded a "grand challenge problem of small-molecule quantum chemistry" \cite{larsson2022chromium}. Unlike most sp molecules, the transition-metal dimers often have spin polarization due to symmetry breaking. 

\begin{figure}
\centering
\includegraphics[width=.8\linewidth]{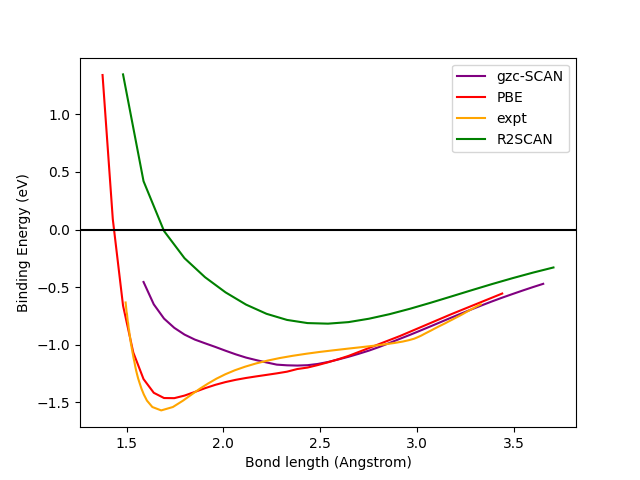}
\caption{Comparison of PBE, r$^{2}$SCAN, and gzc-SCAN PECs for the chromium dimer with the experimental curve \cite{casey1993negative}. SCAN's PEC coincides with that of r$^{2}$SCAN.}

\label{fig:PEC for cr2}
\end{figure}

For the Cr$_{2}$ dimer, while PBE gives a reasonable potential energy curve (PEC), counter-intuitively, SCAN (and r$^{2}$SCAN \cite{furness2020accurate} ) severely underbinds the dimer at all bond lengths. Here, we find that the addition of $\nabla \zeta$ terms to SCAN's correlation resolves the underbinding of the dimer in the tail region of the PEC (Figure \ref{fig:PEC for cr2}), while reducing its underbinding at separations close to the experimental bond length of 1.68 \AA \ \cite{casey1993negative}. This underbinding can be attributed to strong static and dynamic correlations in the dimer, close to its equilibrium separation \cite{larsson2022chromium}, which standard DFAs like PBE and SCAN cannot capture. 

We additionally note that a direct addition of terms in Equation \ref{High-density correction to correlation} to PBE's correlation energy (call gzc-PBE) leads the functional to overbind the Cr$_{2}$ dimer at all bond lengths (PEC presented in the Supplementary Information). This shows that the underbinding of SCAN is not accidental. By underbinding, the functional leaves room for correction by important $\nabla \zeta$-dependent terms in its correlation. Our gzc-PBE calculations were non-self-consistently performed on PBE's orbitals, and the $\nabla \zeta$ terms of Equation \ref{High-density correction to correlation} were left undamped (added as is to PBE's usual correlation). This is not a general-purpose modification of PBE, but a rough estimate of the effect that such a modification would have on Cr$_{2}$.

Unlike in the case of the Cr$_{2}$ dimer, both PBE and r$^{2}$SCAN tend to overbind the Mn$_{2}$ dimer, producing bond lengths that are much smaller than the experimental one \cite{ivanov2021mn,lopez2025revisiting}. Moreover, both functionals incorrectly predict the Mn$_{2}$ dimer to be ferromagnetic (FM) rather than antiferromagnetic(AFM). Here, we evaluate the gzc-SCAN binding energies for both the FM and AFM solutions at their r$^{2}$SCAN equilibrium bond lengths (taken from Ref \cite{lopez2025revisiting}). It was found that gzc-SCAN (0.37 eV) partially reduces the binding energy of r$^{2}$SCAN (0.58 eV). Although gzc-SCAN reduces the energy difference between the FM and AFM solutions (to 0.13 eV), like r$^{2}$SCAN, it predicts an FM ground state. Nonetheless, the fact that gzc-SCAN offers improvements over r$^{2}$SCAN in both underbound (Cr$_{2}$) and overbound (Mn$_{2}$) transition metal dimers, is particularly encouraging.

\section{Conclusions}

In this work, we have carefully added $\nabla \zeta$-dependent terms to SCAN's correlation, ensuring that the modified functional retains most of SCAN's exact constraints. In the high-density limit, the $\nabla \zeta$ terms are recovered in the second-order gradient expansion for the correlation energy. In the low-density limit, the $\nabla \zeta$ terms make SCAN truly independent of spin polarization for a slowly-varying electron gas. The new functional, which we call gradient-zeta-corrected SCAN (gzc-SCAN), satisfies 16 of the 17 exact constraints one can impose on a semi-local functional. The functional has been tested for atoms and molecules, and it has been shown to improve results for transition-metal systems without severely degrading SCAN's excellent performance for $sp$-molecules. 

We mention that while we found gzc-SCAN to give lower magnetic moments than SCAN for ferromagnetic metals, the decrease was only marginal. It has been suggested that SCAN's tendency to overmagnetize metals \cite{fu2018applicability,fu2019density} may stem from the excessive non-locality of its exchange-correlation hole, effectively failing to adequately capture the strong screening effects characteristic of metallic systems.\cite{kaplan2022laplacian}. We find that this problem persists when $\nabla \zeta$-dependent terms are added to SCAN's correlation in the prescribed fashion. In future work, we aim to use $\nabla \zeta$-like terms in the correlation energy densities to develop functionals that provide an improved description of the aforementioned screening effects.

We do not propose the gzc-SCAN functional as a stand-alone functional but do believe that the proposed scheme to add $\nabla \zeta$-dependent terms can make pre-existing meta-GGAs more accurate for transition metal systems. Future work includes testing the effect of adding $\nabla \zeta$-dependent terms to the correlation parts of other functionals and restoring the general Lieb-Oxford bound.

\section{Acknowledgements}

This research was supported in part using high performance computing (HPC) resources and services provided by Information Technology at Tulane University, New Orleans, LA. The work of RM and JPP was supported by the National Science Foundation under grant DMR-2426275. \textcolor{red}{}

\bibliographystyle{unsrt}
\bibliography{bibliography.bib}

\end{document}